\documentclass[%
 reprint,
superscriptaddress,
amsmath,
amssymb,
aps,
pra,
]{revtex4-1}
\usepackage{amssymb}
\usepackage{mathrsfs}
\usepackage{}
\usepackage{color}
\usepackage{graphicx}
\usepackage{dcolumn}
\usepackage{bm}
\usepackage{hyperref}

\begin{document}

\preprint{APS/123-QED}

\title{Spin squeezing via one- and two-axis twisting induced by a single off-resonance stimulated Raman scattering in a cavity}
\author{Gang Liu}%
\affiliation{%
 Department of Physics, Wenzhou University, Zhejiang 325035, China
}%
\author{Ya-Ni Wang}%
\affiliation{%
 Department of Physics, Wenzhou University, Zhejiang 325035, China
}%
\author{Li-Fen Yan}%
\affiliation{%
 Department of Physics, Wenzhou University, Zhejiang 325035, China
}%
\author{Nian-Quan Jiang}%
\affiliation{%
 Department of Physics, Wenzhou University, Zhejiang 325035, China
}%
\author{Wei Xiong}%
\affiliation{%
 Department of Mathematics and Physics, HeFei University, Anhui 230022, China
}%
\author{Ming-Feng Wang}
\email{Corresponding author: mfwang@wzu.edu.cn}
\affiliation{%
 Department of Physics, Wenzhou University, Zhejiang 325035, China
}%


\date{\today}

\begin{abstract}
Squeezed spin states have important applications in quantum metrology and sensing. It has been shown by S{\o}rensen and M{\o}lmer [Phys. Rev. A 66, 022314(2002)] that an effective one-axis-twisting interaction can be realized in a cavity setup via a double off-resonance stimulated Raman scattering, resulting in a noise reduction scaling $\propto 1/N^{2/3}$ with $N$ being the atom number. Here, we show that, by making an appropriate change of the initial input spin state,
it is possible to produce a one-axis-twisting spin squeezing via a \emph{single} off-resonance stimulated Raman scattering, which thus can greatly simplify the realistic implementation. We also show that the one-axis-twisting interaction can be transformed into a more efficient two-axis-twisting interaction by rotating the collective spin while coupling to the cavity, yielding a Heisenberg limited noise reduction $\propto1/N$. Considering the noise effects due to atomic decoherence and cavity decay, we find that substantial squeezing is still attainable with current laboratory techniques.
\begin{description}
\item[PACS numbers]
42.50.Lc, 42.50.Pq, 03.67.Bg, 42.50.Dv
\end{description}

\end{abstract}

\pacs{Valid PACS appear here}
\maketitle


\section{\label{sec:level1}INTRODUCTION}
 Squeezed spin states (SSS) \cite{PRA1993Kitagawa} play essential roles in quantum information processing \cite{JPB2008qf,JPA2014qf,Teles2015} and precision measurement \cite{PRA2007pm,PRL2011pm,PRL2016pm}. They have been shown to have many applications, such as detecting quantum entanglement, improving precision in Ramsey spectroscopy, and making more precise atomic clocks. Recently, various methods have been proposed \cite{PRA2002SS,PRA2010SS,PRA2012LI,NJP2017J-Borregaard,PRA2017Wang,NJP2010ex,PRL2010ex} to create such states, including quantum nondemolition (QND) measurements of collective spins \cite{NJP1998qnd,PRA1999qnd,PRL2000qnd,PRL2000Duan,PRL2016qnd,GSH2015,SSA2018}and nonlinear interaction between spins based on either one-axis twisting (OAT) \cite{PRA1993Kitagawa,PRA2002SS} or two-axis twisting (TAT) \cite{PRA2017Wang,NJP2017J-Borregaard}. Among them, the QND-based methods have the advantage of simple implementation, while, on the other hand, they also suffer the drawback of being difficult to produce highly spin-squeezed state because of inefficient noise-reduction scaling $\propto N^{-1/2}$ with $N$ being the total number of atoms. On the contrary, the nonlinear-interaction-based methods have been proved to work much more efficiently than the QND scheme \cite{PRA1993Kitagawa}, as the
theoretical limit of spin squeezing for OAT scales as $\propto N^{-2/3}$, and the noise reduction for TAT can even reach the Heisenberg limit $\propto N^{-1}$.

To date, much attention has been paid to realize both OAT and TAT evolution in atomic systems. In atomic Bose-Einstein condensates, the OAT Hamiltonian arises from binary atomic collisions \cite{PhysRevLett.107.013601,Gross2010,Riedel2010}. In free-space atomic samples, the interference of
multiple atom-light QND interactions can induce atomic OAT and even TAT interactions \cite{PhysRevLett.94.023003,PhysRevLett.105.193602,PRA2017Wang}. The most studied systems, however, for realizing nonlinear interaction between individual spins, are atomic systems in cavities \cite{PRA2002SS,NJP2017J-Borregaard,PhysRevLett.104.073604,PhysRevA.89.023838,PhysRevLett.110.120402,PhysRevA.86.013828,PhysRevA.77.063811,PhysRevA.75.013804,PhysRevLett.88.243602}. Among them, an impressive work by S\o{}rensen and M\o{}lmer proposed to realize OAT evolution via double off-resonance stimulated Raman scattering (SRS) \cite{PRA2002SS}. They showed that two classical driving fields (which have different central frequency and resonant
Rabi frequency) together with one vacuum cavity mode can simultaneously flip a pair of atoms in a way that is analogous to the emission of correlated photon pair in optical parametric amplification, which creates the entanglement between individual spins, and thus is the essence of OAT squeezing. One of the advantages of this method is that it can produce \emph{unitary} OAT spin squeezing, which, as shown in \cite{1367-2630-20-10-103019},  can yield better clock stability than nonunitary squeezing. Another advantage is that, due to the collective enhancement effect of the atoms-light coupling \cite{PhysRevLett.84.4232,PhysRevLett.86.783}, this method is in principle possible to work in any optical cavity, such as the bad cavity. Recently, this method has also been extended to the case of TAT spin squeezing by Borregaard \emph{et al.} \cite{NJP2017J-Borregaard} via adding another two classical driving fields.

In this paper, we show that by making an appropriate change to the initial input atomic state, it is possible to realize unitary OAT spin squeezing by using a \emph{single} SRS interaction between atoms and light. Our approach inherits all the advantages of \cite{PRA2002SS}. Meanwhile, in
contrast to the mechanism in \cite{PRA2002SS}, our approach requires only a single classical driving field and can
thereby significantly simplify the possible experimental realizations.
By an appropriate coherent control of the collective spin by means of adding a rotation to the spin-polarized direction during
the OAT interaction, we also show that the OAT can be converted into the TAT, leading to faster and stronger squeezing. Compared to the TAT method of \cite{NJP2017J-Borregaard} that uses four classical laser fields, our TAT protocol has the advantages of less experimental resource cost and simpler realization. Further investigation indicates that the present schemes can be made
robust with respect to realistic imperfections, including the atomic decoherence and the cavity decay. During the preparation of this work, we became aware of a proposal \cite{PhysRevA.96.050301} that
is similar to ours. Compared to that work,  pseudo-angular-momentum operators connect to the ground and excited states of the atom, ours opetators connect to the two ground-state sublevels, and therefore our approach has the advantage of long coherence time.

The rest of the paper is structured as follows. In Sec. II, we first review some basic concepts and then give detailed analysis of the processes of
 spin squeezing in an atomic ensemble. In Sec. III, we will consider the noise effects. After that, the experimental
feasibility of the scheme is also discussed. Finally, Sec. IV
contains brief conclusions.

\section{\label{sec:level2}generation of spin squeezing}
\subsection{Ideal case}

\begin{figure}[t]
\centering
\includegraphics[scale=0.45]{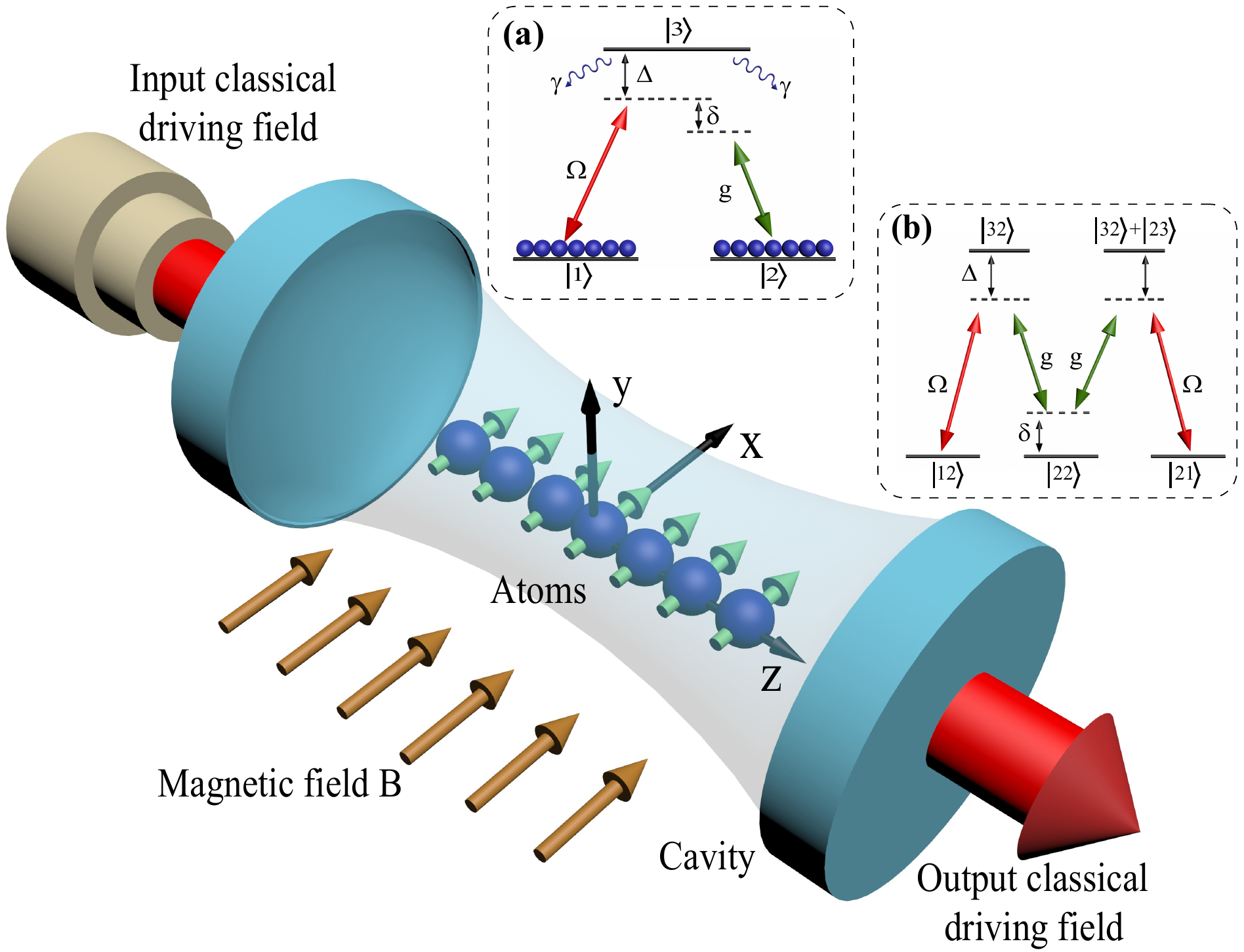}
\caption{ Schematic setup for spin squeezing. A laser beam enters the optical cavity and interacts with an $x$-polarized collective spin to realize the off-resonance coupling between the ground state $|1\rangle$ and the exited state $|3\rangle$. The ground state $|2\rangle$ is coupled to the excited state $|3\rangle$ by a cavity mode which is initially in a vacuum state. Adiabatic elimination of the exited state leads to a nonlinear OAT evolution for the collective spin. Adding a homogeneous magnetic field $B$ to the collective spin along the $x$-direction enables the conversion of OAT into more efficient TAT spin squeezing. (a) Level structure of a single atom. (b) Joint level structure of two atoms. A
double Raman process takes an atom from $|1\rangle$ to $|2\rangle$ and another one from $|2\rangle$ to $|1\rangle$, resulting in the simultaneous flipping of a pair of atoms.}
\label{figure}
\end{figure}
Let us first review the definition of spin squeezing. Consider an atomic system consisting of $N$ two-level atoms, which can be described by the pseudo-angular-momentum operators ${\hat S_z} = \sum\nolimits_k {{{\left( {{{\left| 1 \right\rangle }_k}\left\langle 1 \right| - {{\left| 2 \right\rangle }_k}\left\langle 2 \right|} \right)} \mathord{\left/  {\vphantom {{\left( {{{\left| 1 \right\rangle }_k}\left\langle 1 \right| - {{\left| 2 \right\rangle }_k}\left\langle 2 \right|} \right)} 2}} \right. \kern-\nulldelimiterspace} 2}}$ and $\hat S_+ = \sum\nolimits_k {{{| 1 \rangle }_k}\langle 2 |} $, where the sum is over all the individual atoms, and ${| 1 \rangle }$, ${| 2 \rangle }$ are the two internal states of the atoms. The spin components in three orthogonal directions satisfy the commutation relations $[\hat S_x,\hat  S_y]=i\hbar\hat S_z$, where $\hat S_x=(\hat S_++\hat S_-)/2$ and $\hat S_y=(\hat S_+-\hat S_-)/2i$, resulting in the Heisenberg uncertainty relation $\left(\Delta S_y\right)^2\left(\Delta S_z\right)^2\geq\left|\left\langle\Delta S_x\right\rangle\right|^2/4$. A commonly used measure for the degree of squeezing in an atomic ensemble is Wineland criterion \cite{PRA1992Wineland}, which is defined as
\begin{eqnarray}\label{eq3}
\xi^2=\underset\theta{\text{min}}\left[\frac{N(\Delta\hat S_\theta)^2}{\langle{\hat S}_x\rangle^2}\right],
\end{eqnarray}
where ${\hat S}_\theta=\cos(\theta){\hat S}_y+\sin(\theta){\hat S}_z$ is perpendicular to $\hat S_x$, with $\theta\in [0,2\pi]$. If the squeezing parameter $\textstyle\xi^2<1$, the collective atomic state is said to be spin squeezed.

Our scheme relies on a system of $N$ three-level atoms interacting
with one classical driving field (with Rabi frequency $\Omega$ and frequency $\omega$) and one quantized cavity mode $\hat c$ with frequency $\omega_0$  (see Fig. 1). The cavity mode is initially in a vacuum state, and the atoms each are initially prepared in the equal superposition of their ground states, forming the coherent spin state $|\Psi_{CSS}\rangle=2^{N/2}(|1\rangle+|2\rangle)^{\otimes N}$, which is an eigenstate of the $\hat S_x$ operator with eigenvalue N/2. The classical field is detuned from $1\rightarrow 3$ resonance [with an energy
difference $\omega_{13}$ (hereafter we use the unit $\hbar=1$)] by an amount $\Delta$, while the quantized field is involved in the off-resonant atomic transition $2\rightarrow 3$ (with an energy
difference $\omega_{23}\equiv\omega_{13}$) with detuning $\Delta+\delta$, where $\delta=\omega-\omega_0$ is the two-photon detuing  [see Fig. 1(a)]. For such an atoms-light system, the interaction Hamiltonian can be written as
\begin{eqnarray}
\hat H&=&\omega_0\hat c^\dag c+{\textstyle\sum_k}\omega_{13}{\left| 3 \right\rangle _k}\left\langle 3 \right| \nonumber\\
&+& {\textstyle\sum_k}\left(\frac{\Omega }2{e^{- i\omega t}}{\left| 3 \right\rangle _k}\left\langle 1 \right| +g \hat c{\left| 3 \right\rangle _k}\left\langle 2 \right| +\rm{H.c.}\right),\label{eq2}
\end{eqnarray}
where the first two terms describe the energy of the cavity field and the atoms and the last term accounts for the atoms-light interaction, with the coupling $g=d\sqrt{\omega_0/(2\pi\epsilon_0V_0)}$, where $d$ is the dipole moment of the $|2\rangle\rightarrow|3\rangle$ transition, $\epsilon_0$ is the vacuum
permittivity, and $V_0$ is the mode volume.  Changing to a rotating frame with respect to $\omega_0\hat c^\dag c+{\textstyle\sum_k}\omega{\left| 3 \right\rangle _k}\left\langle 3 \right|$, the interaction Hamiltonian of (\ref{eq2}) is changed into
\begin{align}
\hat H=\Delta{\hat\sigma}_{33}+\frac\Omega2{\hat\sigma}_{31}+g\hat\varepsilon{\hat\sigma}_{32}+\rm{H.c.},\label{eq3}
\end{align}
where we have defined the new operator $\hat\varepsilon=\hat ce^{i\delta t}$ and the collective atomic operators $\hat \sigma_{uv}=\sum_k{\left| u \right\rangle _k}\left\langle v \right|$ with $u,v\in \{1,2,3\}$. Corresponding to this Hamiltonian, one may evaluate the
Heisenberg equations for light and atoms, yielding the following Maxwell-Bloch equations:
\begin{eqnarray}
          \dot {\hat \sigma}_{11} &=& i  \frac \Omega 2 {\hat \sigma}_{31} - i\frac{\Omega^\ast}2{\hat \sigma}_{13},\\
          \dot {\hat \sigma}_{22} &=& ig \hat \varepsilon{\hat \sigma}_{32} - ig^\ast{\hat \sigma}_{23} \hat \varepsilon^\dag, \label{eq5}\\
          \dot {\hat \sigma}_{12} &=& i  \frac \Omega 2 {\hat \sigma}_{32} - ig^\ast{\hat \sigma}_{13} \hat \varepsilon^\dag,\\
           \dot {\hat \varepsilon} &=&-ig^\ast {\hat \sigma}_{23} + i \delta \hat \varepsilon,\label{eq10}\\
          \dot {\hat \sigma}_{13} &=& -i \Delta{\hat \sigma}_{13} - i\frac{\Omega}{2}({ \hat \sigma}_{11} - {\hat \sigma}_{33}) - ig \hat \varepsilon{ \hat \sigma}_{12},\label{eq7}\\
          \dot {\hat \sigma}_{23} &=& -i \Delta{\hat \sigma}_{23} - i\frac{\Omega}{2}\sigma_{21} - ig \hat \varepsilon({ \hat \sigma}_{22} - {\hat \sigma}_{33}),\label{eq8}\\
          \dot {\hat \sigma}_{33} &=& i \frac{\Omega^\ast}{2}\sigma_{13} - i\frac{\Omega}{2}\sigma_{31} + ig^\ast \hat \sigma_{23} \hat \varepsilon^\dag - ig \varepsilon \hat \sigma_{32}.\label{eq9}
         \end{eqnarray}
Next, we assume that (i) the power of the classical driving field is sufficiently
weak and (ii) the detuning is very large $\Delta\gg 1$. With such assumptions, it is reasonable to suppose that the population of the excited state is very small, and thus one may adiabatically eliminate the exited state, leading to $\sigma_{33}\simeq 0$. Large detuning also makes it possible to quickly drive the coherences $\sigma_{13},\sigma_{23}$ into steady states, resulting in
${\hat \sigma _{13}} \simeq  - (\Omega {\hat \sigma _{11}}/2 + g\hat \varepsilon {\hat \sigma _{12}})/\Delta ,{\hat \sigma _{23}} \simeq  - (\Omega {\hat \sigma _{21}}/2 + g\hat \varepsilon {\hat \sigma _{22}})/\Delta $. Furthermore, we also assume that the two-photon detuning is large enough, $\delta\gg 1$, such that there is no significant photon excitation created in the cavity during interaction, which enables the adiabatical elimination of the cavity field, leading to $\hat\varepsilon {\rm{ = }} - {g^*}\Omega {{\hat \sigma }_{21}}/2\Delta \delta$. With these assumptions, the equations for the atomic ground states can be written as
\begin{eqnarray}
{\dot {\hat \sigma} _{12}} =  - i{\kappa _0}{\hat S_z}{\hat \sigma _{12}} - i{\chi _0}{\hat \sigma _{12}},
\dot{\hat\sigma}_{11}=\dot{\hat\sigma}_{22}=0,\label{eq99}
\end{eqnarray}
where we have defined $\kappa_0=|\Omega|^2{|g|^2}/4{\delta\Delta^2}$ and $\chi_0 ={|\Omega|^2}/{4\Delta}$. In the language of pseudo angular momentum operators, Eqs. (\ref{eq99}) can be expressed as
\begin{eqnarray}
{{\dot {\hat S}}_x} &=& {\kappa _0}\left( {{{\hat S}_y}{{\hat S}_z} + {{\hat S}_z}{{\hat S}_y}} +{{\hat S}_y}\right) + {\chi _0}{{\hat S}_y},\label{eq11}\\
{{\dot {\hat S}}_y} &=&  - {\kappa _0}\left( {{{\hat S}_x}{{\hat S}_z} + {{\hat S}_z}{{\hat S}_x}} +{{\hat S}_x}\right) - {\chi _0}{{\hat S}_x},\label{eq12}\\
{{\dot {\hat S}}_z} &=& 0.\label{eq13}
\end{eqnarray}
 From these equations, one may infer that the atomic dynamics are produced by the effective Hamiltonian
\begin{eqnarray}
\hat H_{eff}=-\chi_0\hat S_z-\kappa_0(\hat S_z+\hat S_z^2),\label{eeq14}
\end{eqnarray}
which is exactly the OAT-type interaction \cite{PRA1993Kitagawa}. The first term in (\ref{eeq14}) arises because of ac-Stark shifts of
the ground states, while the rest of the terms stem from the two-photon off-resonance Raman transition. The origin of the nonlinear term in (\ref{eeq14}) can be understood by considering a double Raman process as shown in Fig. 1(b). We assume that an atom in the ground state $|1\rangle$ absorbs a
photon from the classical driving field and emits a photon to the cavity field that is absorbed by another atom in the ground state $|2\rangle$, which then emits back into the classical driving field again, resulting in the effective transitions of the form $|12\rangle\rightarrow|21\rangle$. It is this two-atom process that is kept on resonance and responsible for the spin-spin entanglement (and thus the spin squeezing) generation. The dynamics of such two-atom flipping can be described by an effective
Hamiltonian $\hat S_-\hat S_+=\hat S_x^2+\hat S_y^2=N/2(N/2+1)-\hat S_z^2\propto \hat S_z^2$. It should be mentioned that there exists a probability that the cavity photon emitted by an atom is absorbed by the atom itself, which suppresses the two atom process. This is why we here take the equal-superposition spin state as the input state, as the number of atoms in such a state that participate in cavity-photon reabsorption is around $N/2$, which can greatly suppress the effect of self-reabsorption and therefore makes the two -atom process dominant.

Equation (\ref{eq11}) can be readily solved to yield \cite{PRA1993Kitagawa} ${{\hat S}_x}(t) =\{ {{\hat S}_ + }(0)\exp [i\mu ({{\hat S}_z} + 1/2 + \varphi /\mu )] + \exp [ - i\mu ({{\hat S}_z} + 1/2 + \varphi /\mu )]{{\hat S}_ + }(0)\}/2$, with $\mu = -2\kappa_0 t$ and $\varphi = -\varphi_0 t$ with $\varphi_0=\chi_0+\kappa_0$. Its mean value, after writing the CSS in the basis of Dicke states  $|\Psi_{CSS}{\rangle } = \sum\nolimits_{m =  - S}^S {{{2^{ - S}}\sqrt {(2S)!/[(S + m)!(S - m)!]}}\left| m \right\rangle }$,  can be calculated as: $\langle\Psi_{CSS}|\hat S_x|\Psi_{CSS}\rangle=S \cos^{2S-1}{\frac {\mu}2} \cos{ \varphi }$. For $S\gg 1$ and $|\mu|,|\varphi|\ll1 $, one approximately has $\langle\hat S_x\rangle\simeq S$, which means that almost all the atoms are still polarized along the $x$ direction after the OAT interaction. One thus can use the Holstein-Primakoff approximation \cite{PhysRevA.58.1098} to define new atomic quantum variables ${\hat X_{a}}={{\hat S_{y}}}/{\sqrt{ {S_x}  }},{\hat P_{a}}={{\hat S_{z}}}/{\sqrt{{S_x}}}$, which satisfy $[ {\hat X_{a}},{\hat P_{a}} ]=i$ and have zero
mean $\langle\hat X_{a}\rangle=\langle\hat P_{a}\rangle=0$ and a normalized variance $(\Delta\hat X_{a})^2=(\Delta\hat P_{a})^2=1/2$ for the initial CSS. In this language, the solutions to Eqs. (\ref{eq12}) and (\ref{eq13}) can be expressed as
\begin{eqnarray}
\hat X_a^{\rm out} = \hat X_a^{\rm in} + \alpha \hat P_a^{\rm in} + \beta ,
\hat P_a^{\rm out} = \hat P_a^{\rm in},\label{eq14}
\end{eqnarray}
where ``in" and ``out" refers to the atoms before and after the interaction and we have defined the coupling constant $\alpha=S\mu$ and the displacement parameter $\beta=\sqrt{S}\varphi$ that arises because of the linear term of the Hamiltonian (\ref{eeq14}). Apparently, the spin state of (\ref{eq14}) is produced by first squeezing the spin state via a Hamiltonian quadratic in $\hat P_a^2$, and then displacing the SSS in the phase space along the $\hat P_a$ direction by an amount $\beta$. Since the displacement operation (linear operation) in phase space does not reduce the atom-atom entanglement created by the OAT evolution \cite{RevModPhys.77.513}, the displacement $\beta$ can then be neglected when we estimate the amount of squeezing of the atomic system. To see how much squeezing is created, we rotate the spin state around the $x$-axis by the unitary transformation $\hat X_\theta^{\rm out}=\exp(i\theta\hat H_{SR}) \hat X_a^{\rm out}\exp(-i\theta\hat H_{SR})=\cos{\theta} \hat X_a^{\rm in}+(\alpha\cos{\theta}+\sin{\theta})\hat P_a^{\rm in}$, where $\hat H_{SR}=-2\hat S_x\simeq \hat X_a^2+\hat P_a^2$ is the spin-rotation Hamiltonian \cite{PhysRevLett.91.060401}. Optimizing the variance $(\Delta\hat X_\theta^{\rm out})^2$ with respect to $\theta$,  we finally get $\xi_{OAT}^2=2(\Delta\hat X_\theta^{\rm out})^2=1+\alpha^2/2-(\alpha^4/4+\alpha^2)^{1/2}\Rightarrow \mathop{\lim}\limits_{\alpha\rightarrow \infty}1/\alpha^2$, for $\theta=\arctan (2/\alpha)/2+\pi/2$.

The amount of squeezing can be dramatically increased if one can transform the OAT into the TAT \cite{PhysRevLett.107.013601}. To do so, we add a rotation about the $x$-direction during OAT interaction with an angular frequency $\Omega_0$ \cite{PhysRevA.92.063610,PhysRevA.87.051801,PhysRevA.91.053826} (which can be realized by applying a homogeneous magnetic field along the $x$ axis as shown in Fig. 1), resulting in the Hamiltionian $\hat H_{TAT}=\Omega_0\hat H_{SR}/2+\hat H_{eff}=\Omega_0(\hat X_a^2+\hat P_a^2)/2-\kappa_0 S\hat P_a^2-\sqrt{S}\phi_0\hat P_a=\kappa_0S(\hat X_a^2-\hat P_a^2)/2-\sqrt{S}\phi_0\hat P_a$ for $\Omega_0=\kappa_0S$, which is exactly the TAT-type interaction \cite{PRA1993Kitagawa} and squeezes the spin fluctuations at a rate that scales exponentially with coupling constant, that is $\xi^2_{TAT}=\exp(-\alpha)$. In contrast to the OAT that creates squeezing polynomially, the exponential scaling of the TAT method will greatly enhance the entanglement between individual atoms and thus enable us to perform nontrivial control of collective spin.

\subsection{Noise effect}
So far, we have neglected the noise effects. As in reality,
the photons leak out from the cavity into the environment at a rate $\kappa$ and the excited state decays to the ground state with a radiative decay rate $\gamma_{13}=\gamma_{23}\equiv \gamma=\omega_0^2d^2/(3\pi\epsilon_0c^3)$ \cite{book11}. In the presence of decays as well as spin rotation about the $x$ direction,  the time evolution of the atomic operators (see the appendix for more details) can be written as
\begin{eqnarray}
\frac{d}{{dt}}\left( {\begin{array}{*{20}{l}}
{{{\hat X}_a}}\\
{{{\hat P}_a}}
\end{array}} \right) = \mathcal{G}\left( {\begin{array}{*{20}{l}}
{{{\hat X}_a}}\\
{{{\hat P}_a}}
\end{array}} \right) - \sqrt S \left( {\begin{array}{*{20}{l}}
{\varphi_0}\\
\eta
\end{array}} \right)
+ \sqrt {2\eta } \left( {\begin{array}{*{20}{l}}
{{{\hat {\mathcal{F}}}_y}}\\
{{{\hat {\mathcal{F}}}_z}}
\end{array}} \right)\nonumber\label{eq15}
\end{eqnarray}
for the case of $r_0=\kappa/2\delta\ll 1$, with
\begin{eqnarray}
\mathcal{G}={\Omega _0}\left( {\begin{array}{*{20}{c}}
0&{ 1}\\
-1&0
\end{array}} \right) - 2S{\kappa _0}\left( {\begin{array}{*{20}{c}}
0&1\\
0&0
\end{array}} \right) - \eta \left( {\begin{array}{*{20}{c}}
1&0\\
0&1
\end{array}} \right)\nonumber,
\end{eqnarray}
where $\eta=\chi_0\gamma/\Delta$ is the atomic decay parameter and $\hat {\mathcal{F}}_y,\hat {\mathcal{F}}_z$  are Langevin noise operators that have zero means and satisfy $\langle\hat{\mathcal{F}}_y(t) \hat{\mathcal{F}}_z(t')\rangle= i\delta(t-t')/2$, and $\langle\hat{\mathcal{F}}_y(t)\hat{\mathcal{F}}_y(t')\rangle=\langle\hat{\mathcal{F}}_z(t)\hat{\mathcal{F}}_z(t')\rangle=\delta(t-t')/2$. The first term of $\mathcal{G}$ represents atoms turn with $\Omega_0$ around the $x$ axis. The second term in $\mathcal{G}$ denotes the coherent OAT interaction induced by light, while the third term stands for the transverse decay of atoms caused by optical pumping. Note that $\hat P_a$ is now also displaced at a rate proportional to $\eta$, which is due to the ground-state population transfer from 1 to 2 induced by the strong light field. The solution to this differential equation is
\begin{eqnarray}
\left( {\begin{array}{*{20}{l}}
{\hat X_a^{\rm out}}\\
{\hat P_a^{\rm out}}
\end{array}} \right) &=& \mathcal{A}\left( t \right)\left( {\begin{array}{*{20}{l}}
{\hat X_a^{\rm in}}\\
{\hat P_a^{\rm in}}
\end{array}} \right) - \mathcal{A}\left( t \right)\int_0^t {d\tau {\mathcal{A}^{ - 1}}\left( \tau  \right)} \nonumber\\
 &&\times \left[ {\sqrt S \left( {\begin{array}{*{20}{l}}
{{\varphi _0}}\\
\eta
\end{array}} \right) - \sqrt {2\eta } \left( {\begin{array}{*{20}{l}}
{{{\hat {\mathcal{F}}}_y\left( \tau \right)}}\\
{{{\hat {\mathcal{F}}}_z\left( \tau \right)}}
\end{array}} \right)} \right],\label{eq16}
\end{eqnarray}
with the homogeneous solution $\mathcal{A}\left( t \right)=\exp(\mathcal{G}t)$. For the case of OAT squeezing (that is, $\Omega_0=0$), we have $\mathcal{A}(t)={e^{ - \eta t}}\left( {\begin{array}{*{20}{c}}
1&\alpha \\
0&1
\end{array}} \right)$ and obtain directly from (\ref{eq16})
\begin{eqnarray}
\hat X_a^{\rm out} &=& \sqrt {1 - {\eta _0}} \left( {\hat X_a^{\rm in} + \alpha \hat P_a^{in} + \beta' } \right) + \sqrt {{\eta _0}} {{\hat{\mathcal{F}}}_X},\label{eq116}\\
\hat P_a^{\rm out} &=& \sqrt {1 - {\eta _0}} \left( {\hat P_a^{\rm in} + \beta ''} \right) + \sqrt {{\eta _0}} {{\hat{{\mathcal{F}}}}_P},\label{eq117}
\end{eqnarray}
\begin{figure*}
[t]\resizebox{17.5cm}{!}
{\includegraphics[scale=1.3]{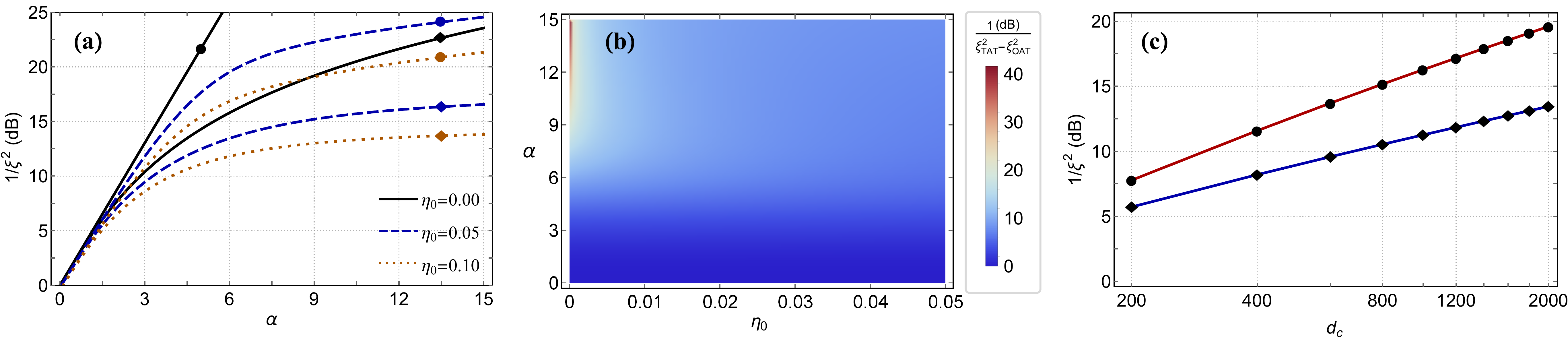}}
\caption{\label{fig:figure3} (Color online) (a) Performance of OAT (lines with diamond) and TAT (lines with circle) protocols varies with coupling strength $\alpha$ for various values of atomic decay. (b) Squeezing difference of the two proposed squeezing protocols vs coupling strength $\alpha$ and atomic decay $\eta_0$. (c) The achievable squeezing of OAT (line with diamonds) and TAT (line with circles) vs cavity OD $d_c$ for $r_0=0.1$. }
\end{figure*}
where we have defined the parameters $\eta_0=2\eta t, \beta'=\beta+\eta_0\alpha/4$, and $\beta''=\sqrt{S}\eta_0/2$, and used the condition $\eta_0\ll 1$. The modified noise operators are of the form: $\hat{\mathcal{F}}_X=\frac{1}{\sqrt t}e^{-\eta t}\int_0^td\tau e^{\eta \tau}[\hat{\mathcal{F}}_y(\tau)-\alpha(\tau-t)/t\hat{\mathcal{F}}_z(\tau)],\hat{\mathcal{F}}_P=\frac{1}{\sqrt t}e^{-\eta t}\int_0^td\tau e^{\eta \tau}\hat{\mathcal{F}}_z(\tau)$, which can be easily checked to have $\langle\hat{\mathcal{F}}_X\rangle=\langle \hat{\mathcal{F}}_P\rangle=0,\langle\hat{\mathcal{F}}_X \hat{\mathcal{F}}_P\rangle\simeq i(1-i\alpha/2)/2,\langle\hat{\mathcal{F}}_X^2\rangle\simeq (1+\alpha^2/3)/2$, and $\langle \hat{\mathcal{F}}_P^2\rangle\simeq 1/2$. With help of the atomic input-output relations (\ref{eq116}) and (\ref{eq117}), one may calculate the variance $(\Delta\hat X_\theta^{\rm out})^2$, and thus obtain the optimized variance
\begin{eqnarray}
2(\Delta\hat X_\theta^{\rm out})^2 &=& 1 + \frac{{{\alpha ^2}}}{2}\left( {1 - \frac{2}{3}{\eta _0}} \right)\nonumber\\
 &&- \sqrt {{{\left( {1 - \frac{2}{3}{\eta _0}} \right)}}^2\frac{{{\alpha ^4}}}{4} + {{\left( {1 - \frac{1}{2}{\eta _0}} \right)}}^2{\alpha ^2}}\nonumber\\
 &&\Rightarrow \frac{1}{\alpha^2}+\frac{\eta_0}{3},\quad\quad\quad\quad\quad\alpha\gg 1,\label{eq1118}
\end{eqnarray}
for $\theta=\arctan [(2-\eta_0)/(\alpha-2\eta_0\alpha/3)]/2+\pi/2$. Equation (\ref{eq1118}) shows that the atomic decay sets a limit, that is, $\eta_0/3$, to the highest degree of squeezing that can be achieved. For atoms situated on the cavity antinode, the coupling $g$ can be conveniently expressed in terms of the excited-state linewidth  \cite{phd11} $|g|^2=\gamma\kappa d_c/(4N)$, with the cavity optical depth (OD) $d_c=\frac{2\mathcal {F}}{\pi}d_0=\frac{2\mathcal {F}}{\pi}\frac{N\sigma_0}{A_0}$, where $\mathcal {F}$ is the cavity finesse, $d_0=N\sigma_0/A_0$ is the sample's OD in free space, $\sigma_0$ is the photon-absorption cross
section of an atom, and $A_0$ is the effective cross-sectional area of the antinode. The coupling constant $\alpha$ can then be re-expressed as $\alpha=r_0 d_c\eta_0/2$. Consequently, the amount of squeezing for large $\alpha$ may be written as $2(\Delta\hat X_\theta^{out})^2= 1/(r_0 d_c\eta_0/2)^2+\eta_0/3\geq 3^{1/3}/(r_0 d_c)^{2/3}\propto 1/N^{2/3}$, which is exactly the OAT scaling as mentioned above.

 For the case of TAT squeezing (that is, $\Omega_0=S\kappa_0$), we have $\mathcal{A}(t)=e^{-\eta t}\left( {\begin{array}{*{20}{c}}
\cosh\frac{\alpha}{2}&-\sinh\frac{\alpha}{2} \\
-\sinh\frac{\alpha}{2}&\cosh\frac{\alpha}{2}
\end{array}} \right)$, and thus one may derive the input-output relations for the atomic quadrature $\hat X_{\pi/4}$ from Eq. (\ref{eq16})
\begin{eqnarray}
\hat X_{\frac{\pi }{4}}^{\rm out}{\rm{ }} = \sqrt {1 - {\eta _0}} \left[ {{e^{ - \frac{\alpha }{2}}}\hat X_{\frac{\pi }{4}}^{\rm in} + \frac{{ {\beta  - \beta ''} }}{{\sqrt 2 }}} \right] + \sqrt {{\eta _0}} {\hat {\mathcal{F}}_{\frac{\pi }{4}}},
\end{eqnarray}
with ${\hat {\mathcal{F}}_{\frac{\pi }{4}}}=\frac{1}{\sqrt{2t}}e^{-(\eta_0+\alpha)/2}\int^t_0d\tau e^{(\eta+S\kappa_0)\tau}[{\hat {\mathcal{F}}_{z}}(\tau)-{\hat {\mathcal{F}}_{y}}(\tau)]$. Its variance can be directly calculated to yield
\begin{eqnarray}
2(\Delta\hat X_{\frac{\pi }{4}}^{\rm out})^2&=&\left(1-\eta_0\right)e^{-\alpha}+\frac{\eta_0\left[1-(1-\eta_0)e^{-\alpha}\right]}{\eta_0+\alpha}\nonumber\\&&\Rightarrow \frac{\eta_0}{\alpha}\propto\frac{1}{N},\quad\quad\quad\quad\quad\alpha\gg 1,\label{eq118}
\end{eqnarray}
which indicates that the TAT scheme produces a Heisenberg-scaling squeezing. Furthermore, it is also required to take into account the effect of  the $x$-component decay according to the definition given in Eq. (\ref{eq3}), which is ${\langle {\hat S_x}\rangle ^2} \simeq (1 - {\eta _0})N/2$ [as can be derived from Eq. (\ref{a2})]. Finally, we are able to calculate the squeezing parameter $\xi^2\simeq 2(\Delta\hat X_\theta^{\rm out})^2/(1 - {\eta _0})$ and plot in Fig. 2(a) the amount of squeezing in their dependence on coupling strength $\alpha$ for various
values of atomic decay. As can be seen from the figure that, if the atomic decay is small than $10\%$, a high degree of squeezing
 larger than 10 dB would be obtainable for the interaction parameter $\alpha=5$. Besides, as expected, the TAT scheme works much more efficiently than the OAT scheme even in the presence of noises. While a further investigation of the performance of the protocols versus atomic decay [as shown in Fig. 2(b)] indicates that the TAT protocol is sensitive to noises, the OAT protocol, on the contrary, is much more robust to the atomic decay. Consequently, an extremely low decay rate is required to fully preserve the advantages of the TAT protocol.
 In a realistic implementation, it is quite convenient to use the accessible experimental parameter OD to assess the performance of the proposed protocols. In Fig. 4(c), the best achievable squeezing (optimized with respect to $\eta_0$) of the two proposed protocols versus cavity OD $d_c$ is also plotted. For room-temperature vapors whose OD in free space is around 30 \cite{JPB2008qf}, if one set the parameter $r_0=0.1$ and the finesse $\mathcal {F}\simeq100$, degrees of squeezing created by OAT and TAT should be as high as 13.4 and 19.6 dB, respectively.

Giving an estimation of the relevant parameters is helpful for implementing realistic experiments.
 We consider an atomic sample containing $5\times 10^6$ atoms and chose a realistic cavity coupling parameter $g=(2\pi)100$ kHz. If one chooses the parameters $\gamma=\kappa\sim 10^2g,\Omega\sim 10^4g,\Delta\sim 10^5g$, and $\delta\sim 5\times 10^2 g$,  $\alpha\simeq 5$ is obtainable for interaction time $t$ around $0.3~\mu s$, and at the same time, we have $\eta_0< 10\%,r_0\ll 1$. With these settings, one is to obtain the amount of squeezing larger than 10 dB.

\section{Conclusion}
In conclusion, we have presented a realistic
scheme for generating highly spin-squeezed state of an atomic ensemble in an optical cavity. The process is based on
off-resonance SRS interaction between light and spin-polarized atomic ensembles. By sending a strong pulse through polarized atomic vapors placed in an optical cavity that is initially in a vacuum state, we find that unitary OAT squeezing can be realized. As the interaction between cavity field and atoms increases, so does the
higher the degree of squeezing. We also show that the OAT protocol can be transformed into the more efficient TAT protocol by just adding a homogeneous magnetic field along the spin-polarized direction. The proposed schemes are also tested by adding different noise effects, and we found that (i) substantial squeezing off more than 10 dB is still obtainable even in the presence of $10\%$ atomic decay, (ii) although the performance of the TAT protocol is, in general, superior to the OAT protocol, it is much more sensitive to the atomic decay, and (iii) the OAT protocol is quite robust against noises. We thus believe that, although the OAT protocol is not superior to  the TAT protocol in squeezing scaling, its good characteristics of easily surviving in a noisy environment as well as simpler experimental setup make it applicable to a wide range of atomic systems.
We expect that the proposed protocols can be beneficial in the context of quantum information processing and quantum metrology.
\begin{acknowledgments}
We thank Yanhong Xiao for helpful discussions. We acknowledge funding from the National Natural Science Foundation of China
(NNSFC) under Grants No. 11504273 and No. 11804074.
\end{acknowledgments}
\appendix
\section{Details of the derivation of the equations of motion with noises
}

In this appendix, we analyze the performance of our proposal
in the presence of spontaneous emission and cavity decay. By taking into account the noise effects, the Maxwell-Bloch equations of (\ref{eq5})-(\ref{eq9}) are then changed into \cite{PhysRevA.76.033804}
\begin{eqnarray}
	\dot {\hat \sigma}_{11}  &=& -\Omega_0 \hat S_y + i   \frac \Omega2{ \hat \sigma}_{31} - i \frac{ \Omega^\ast}{2}{ \hat \sigma}_{13} + \gamma{ \hat \sigma}_{33}+ F_{11},\label{25}\nonumber\\
	\dot {\hat \sigma}_{22} &=&\Omega_0 \hat S_y + i g \hat \varepsilon{ \hat \sigma}_{32} - ig^\ast{ \hat \sigma}_{23} \hat \varepsilon^\dag + \gamma{ \hat \sigma}_{33}+ F_{22},\label{26}\nonumber\\
	\dot {\hat \sigma}_{12} &=&  i  \Omega_0  \hat S_z + i   \frac{\Omega}{2}{ \hat \sigma}_{32} - ig^\ast{ \hat \sigma}_{13} \hat \varepsilon^\dag,\label{27}\nonumber\\
	\dot {\hat\varepsilon} &=& - \frac\kappa2\hat\varepsilon - ig^\ast{\hat\sigma}_{23} + \sqrt\kappa{\hat\varepsilon}_{in}+i\delta\hat\varepsilon,\nonumber\\
	\dot {\hat \sigma}_{13} &=& -i\frac{ \Omega_0}2 \sigma_{23} - (i  \Delta + \gamma){ \hat \sigma}_{13}-i \frac\Omega2({ \hat \sigma}_{11}-{ \hat \sigma}_{33})\nonumber\\
	                          &&- ig\hat\varepsilon{\hat\sigma}_{12} + F_{13},\nonumber\\
	\dot {\hat \sigma}_{23} &=& -i\frac{ \Omega_0}2 \sigma_{13} - (i\Delta+\gamma){\hat\sigma}_{23} - ig\hat\varepsilon({\hat\sigma}_{22}-{\hat\sigma}_{33})\nonumber\\
                       	&&- i\frac\Omega2{\hat\sigma}_{21}+F_{23},\nonumber\\
	\dot {\hat \sigma}_{33} &=&i\frac{\Omega^\ast}2{\hat\sigma}_{13} - i\frac\Omega2{\hat\sigma}_{31}+ig^\ast{\hat\sigma}_{23}\hat\varepsilon^\dag
	                          - ig\hat\varepsilon{\hat\sigma}_{32} \nonumber\\
&&- 2\gamma{\hat\sigma}_{33}+F_{33}, \label{a1}
\end{eqnarray}
where we have introduced the radiative decay rate of the excited state $|3\rangle$, $\gamma_3=\gamma_{13}+\gamma_{23}=2\gamma$ (we assume $\gamma_{13}=\gamma_{23}\equiv \gamma$), the Langevin noise operators $F_{uv}$ for the atomic operators, the cavity decay rate $\kappa$, and the input field $\hat \varepsilon^{in}$ for the cavity mode. The correlation functions of Langevin noise operators can be derived by using the the generalized Einstein relation \cite{PhysRevA.76.033804,JOB} $\langle \hat F_{uv}(t)\hat F_{u'v'}(t')\rangle=\langle\mathcal {D}(\hat\sigma_{uv}\hat\sigma_{u'v'})-\mathcal {D}(\hat\sigma_{uv})\hat\sigma_{u'v'}-\hat\sigma_{uv}\mathcal {D}(\hat\sigma_{u'v'})\rangle\delta(t-t')$, where $\mathcal {D}(\hat\sigma_{uv})$ denotes the evolution for ${\hat\sigma}_{uv}$ obtained from
the Heisenberg-Langevin equation but with the Langevin noise omitted. The input cavity satisfies $[\hat\varepsilon_{in}(t),\hat\varepsilon_{in}^\dag(t')]=\delta(t-t')$.
Here we assume without loss of generality that there is no decay between 1 and 2 (as the coherence time of the ground state is normally much longer than the interaction time $t$ in a realistic implementation). Besides, we also introduced a spin rotation to the system about the $x$ direction, which is obtained by adding to the Hamiltionian of Eq. (\ref{eq3}) a time-independent term $\Omega_0 \hat J_x$, resulting in the terms in Eqs. (\ref{a1}) that are proportional to $\Omega_0$. Corresponding to this set of coupled equations, the evolution for the ground states can be derived along
the lines outlined in the main context above to give
\begin{eqnarray}
       \dot {\hat \sigma}_{11} &=& -\Omega_0 {\hat S}_y - \eta{\hat\sigma}_{11} +\hat{\mathcal{F}}_{11},  \nonumber\\
       \dot {\hat \sigma}_{22} &=& \Omega_0 {\hat S}_y +\eta{\hat\sigma}_{11}+\hat{\mathcal{F}}_{22},
                                  \nonumber\\
       \dot {\hat \sigma}_{12} &=& i\Omega_0 {\hat S}_z - \left(\frac{|\Omega|^2}{4\Delta_\gamma^\ast}
                                  +  \frac{i|\Omega|^2|g|^2}{2\delta_{\kappa/2}^\ast\Delta\Delta_\gamma^\ast}S_z\right){\hat\sigma}_{12}+\hat{\mathcal{F}}_{12},\label{a2} \nonumber\\
\end{eqnarray}
where $\eta=\chi_0\gamma/\Delta$ is an optical pumping rate and we have defined
$\delta_{\kappa/2}=\frac\kappa2-i\delta,\Delta_\gamma=\gamma+i\Delta$, and $\hat{\mathcal{F}}_{11},\hat{\mathcal{F}}_{22},\hat{\mathcal{F}}_{12}$ are modified Langevin noise operators.
 In deriving Eqs. (\ref{a2}) we have assumed the angular frequency $\Omega_0\ll \Delta$ and thus neglected its influence on the adiabatic-elimination procedure. From Eqs. (\ref{a2}) one may directly deduce the time evolution of the collective spin operators
\begin{eqnarray}
           {{\dot {\hat S}}_y} &=&   {\Omega _0}{{\hat S}_z} - \frac{\kappa_0}{1+r_0^2}\left( {{{\hat S}_x}{{\hat S}_z} + {{\hat S}_z}{{\hat S}_x} + {{\hat S}_x}} \right)\nonumber\\
 &&+ \frac{r_0\kappa_0}{1+r_0^2}\left( {{{\hat S}_y}{{\hat S}_z} + {{\hat S}_z}{{\hat S}_y} + {{\hat S}_y}} \right)\nonumber\\
 &&- {\chi _0}{{\hat S}_x} - \eta {{\hat S}_y} + \sqrt {2S\eta } {\hat {\mathcal{F}}_y}, \label{a4}\\ 	
           \dot {\hat S}_z&=&  -\Omega_0 {\hat S}_y - S\eta- \eta{\hat S}_z  +\sqrt{2S\eta}\hat{\mathcal{F}}_z,\label{a5}
\end{eqnarray}
where $r_0=\kappa/2\delta$ and
 we have neglected the ac-Stark shifts of
the ground states induced by the cavity mode. We also used the relation $\hat\sigma_{11}\simeq\hat S_z+S$ in the right-hand side of Eq. (\ref{a5}),
and defined the new vacuum noise operators $\hat{\mathcal{F}}_y= (\hat{\mathcal{F}}_{12}-\hat{\mathcal{F}}_{12}^\dag)/2\sqrt{S\eta}i$, $\hat{\mathcal{F}}_z= (\hat{\mathcal{F}}_{11}-\hat{\mathcal{F}}_{22})/2\sqrt{S\eta}$, which, according to the Einstein relations,  have the correlations $\langle\hat{\mathcal{F}}_y(t) \hat{\mathcal{F}}_z(t')\rangle\simeq i\delta(t-t')/2$ and $\langle\hat{\mathcal{F}}_y(t)\hat{\mathcal{F}}_y(t')\rangle=\langle\hat{\mathcal{F}}_z(t)\hat{\mathcal{F}}_z(t')\rangle\simeq \delta(t-t')/2$. The above equations indicate that the noises cause a decay of the transverse spin components and a redistribution of the populations of the ground states [since $\langle \hat S_z\rangle\propto S\eta$, as can be seen from Eq. (\ref{a5})]. Note that the second line of Eq. (\ref{a4}) arising because of cavity decay is negligible in the limit of $r_0\ll 1$, which is the case in the main context.

\bibliographystyle{unsrt}%
\bibliography{ref}

\end{document}